\documentclass[sigconf]{acmart}

\usepackage{booktabs} % For formal tables
\usepackage[]{todonotes}
\usepackage[]{algorithm2e}
\usepackage{listings}

\setcopyright{rightsretained}
\acmDOI{10.475/123_4}

\acmISBN{123-4567-24-567/08/06}

\acmConference[RecSys'18]{Twelfth ACM Conference on Recommender Systems}{October 2018}{Vancouver, Canada}
\acmYear{2018}
\copyrightyear{2018}

\acmArticle{4}
\acmPrice{15.00}

\usepackage{color}

\begin{document}
\title{Attainment Ratings for Graph-Query Recommendation}
%\titlenote{Produces the permission block, and
%  copyright information}
%\subtitle{Extended Abstract}
%\subtitlenote{The full version of the author's guide is available as
%  \texttt{acmart.pdf} document}

%AUTHORSHIP CURRENTLY COMMENTED OUT
\author{Hal Cooper}
\affiliation{%
  \institution{Columbia University}
  \streetaddress{500 W 120th Street}
  \city{New York}
  \state{New York}
  \country{USA}
  \postcode{10027}
}
\email{hal.cooper@columbia.edu}

\author{Garud Iyengar}
\affiliation{%
  \institution{Columbia University}
  \streetaddress{500 W 120th Street}
\city{New York}
\state{New York}
\country{USA}
\postcode{10027}
}
\email{garud@ieor.columbia.edu}

\author{Ching-Yung Lin}
\affiliation{%
  \institution{Graphen, Inc.}
  \streetaddress{500 5th Avenue, Floor 46}
  \city{New York}
  \state{New York}
  \country{USA}
  \postcode{10110}
  }
\email{cylin@graphen.ai}
%
%% The default list of authors is too long for headers.

\begin{abstract}
The video game industry is larger than both the film and music
industries combined. Recommender systems for video games have
received relatively scant academic attention, despite the uniqueness of the medium and its data. In this paper, we introduce
a graph-based recommender system that makes use of interactivity, arguably
the most significant feature of video gaming. We show that the use of
implicit data that tracks user-game interactions and levels of attainment
(e.g. Sony Playstation Trophies, Microsoft Xbox Achievements) has high
predictive value when making recommendations. Furthermore, we argue that
the characteristics of the video gaming hobby (low cost, high duration,
socially relevant) make clear the necessity of personalized, individual
recommendations that can incorporate social networking information. We
demonstrate the natural suitability of graph-query based recommendation
for this purpose.  
\end{abstract}





\maketitle

\section{Introduction}
Steam \cite{Steam2003} is the largest digital distribution service for video games on
the PC, Mac, and Linux platforms, %  It is the largest such service currently
% existing,
with more than 18,000 available video game products. Products on
Steam are actively marketed to (more than 100 million) users, with regular discounted sales
events and pop-ups of recommended products. 

Video distributions services such as Netflix \cite{Gomez-Uribe2015} and Amazon
\cite{Leino2007} have long used recommender systems to help customers
become aware of suggested products. 
% On its face, this seems like a recommendation problem similar to that of
% many online retailers, such as (perhaps the most visible users of product recommendation
% systems). 
Netflix is the runner of the well-known ``Netflix Prize''
\cite{Netflix2009} for recommendation systems using user-movie rating
triplets. These services % heavily
primarily
employ % the use of so-called
``explicit''
ratings data, wherein a user directly and deliberately inputs some form of
rating for a product. % ) is the most common method used for product
% recommendation.
% mplicit data (data recorded
In this context, implicit data, i.e. data that may \emph{implicitly} indicate
user preferences, but that does not involve the user explicitly designating
a score or writing a review, is often seen as less causally informative,
more biased, and less precise. Unary ratings (where it is possible to
``like'' a product e.g. click on an ad, but not possible to register
dislike), are common in implicit rating scenarios \cite{Hu2008} and
exemplify many of the aforementioned issues. Though Steam does allow users
to post reviews of purchased products, the use of such data in making
recommendations could be called into question due to review bombing
\cite{Kuchera2017}. Fortunately, there exists a wealth of additional data
less susceptible to manipulation.  

Steam is also a software platform on which
purchased products are managed, organized, and \textit{run}. It is the
interactive nature of video games that distinguishes games from other forms
of media. Rather than a passive process of consumption, playing video games
involves making conscious choices about how, when, and why to perform
certain interactions. It is this interactive experience of the consumers with
the product that they purchase (and other users who have made the same purchase) that 
% Implicit data
differentiates digital distribution systems like Steam 
from % online
retailers like Netflix and Amazon that % (primarily) 
sell  physical and digital 
products through an online marketplace. Furthermore, in addition to being
a marketplace, Steam 
 is both an active online community with a
well-defined social network.
In this paper, we argue that the choices players'
make % regarding these interactions 
are indicative of their preference of
titles, and can therefore be used for the purposes of product
recommendation. Furthermore, we argue that presently recorded data
directly reflects these actions.  
 
The boom in multiplayer online video games % online became prominent during the six
% and seventh generation of videogame consoles,
occurred around the same time as the increase in the popularity of 
% the use of massive
online
social networks % was also experiencing a boom 
\cite{Edosomwan2011}. Perhaps
in response, Microsoft introduced the Xbox Live system
\cite{Microsoft2002}. In addition to centralizing and streamlining online
play, Xbox Live was created as a social network. Users had avatars that
represented their likenesses, and added other users, known in real-life,
or encountered during online play, to their list of friends through the use of
unique ``GamerTags''. The concept of an ``Xbox Live Achievement''
(hereafter simply referred to as an ``achievement'') was created,
``gamifying'' the play of video games \cite{Jakobsson2011}. Achievements
are  awarded when players completed certain (potentially difficult)
tasks in a game, and these then contribute to a  ``GamerScore''. This
GamerScore, as well as the achievements comprising it, would be publicly
visible, giving users a form of bragging rights that demonstrated their
skill. This ``gamification'' of playing video games (that is, giving the
process of playing and completing video games the property of a game at a
meta-level) served to entice players to become more competitive (thereby
participating in more play and being more invested in the Xbox Live social
network) and makeadditional purchases (with so called ``play-to-win''
models now being a common method for monetization of game products
\cite{McKinney2017}). 
 
Digital marketplaces for other consoles \cite{Niizumi2006} and computers
(e.g. Steam) soon followed suit. Through the use of the Steam launcher or
website, Steam members can view the products owned by other members, as
well as their Steam Achievements. Users can also view the global stats of
achievements, such as the percentage of owners of a game who have gained a
particular achievement, as well as the requirements for gaining it (though
some particularly difficult achievements, or achievements that spoil the
game, may have their details hidden). This data is also available through
the Steam WebAPI \cite{Steam2016}. Prior to April 10, 2018, the vast
majority of this data was publicly accessible. After this date, most of
the data was made private \cite{Bratt2018} through the retroactive
replacement of a user opt-out procedure with an opt-in procedure. 

\section{Attainment Ratings}
In this paper we argue that achievements are highly informative
implicit data that can be used to determine a score representing
how much a particular user likes a particular game. The rationale is quite
simple; the more a user enjoys a game, the more they will attempt to
complete everything it has to offer. We compute an overall achievement
score for a particular user by combining each individual achievement
weighted by difficulty level computed using 
% Through the use of examining the
% achievements that a particular user has gained for a game, we approximate
% the level of achievement. Furthermore, by using 
global achievement rates.
% we can determine how difficult each achievement is to obtain.  

Hours played has been proposed in the literature \cite{Hamari2011} as a metric  to
approximate user preferences. 
% for instance, in 
% use ``hours played'' is used as a 
% metric.
However, there are significant issues with the use of simple play
times. For instance, the length of games is not uniform. Certain genres
(e.g. Role-Playing games) are typically significantly longer in length
than other genres (such as Action games), and many highly
acclaimed games are extremely short (such as the Playstation title,
Journey). 

\begin{figure}
	\centering
	\includegraphics[width=1.0\linewidth]{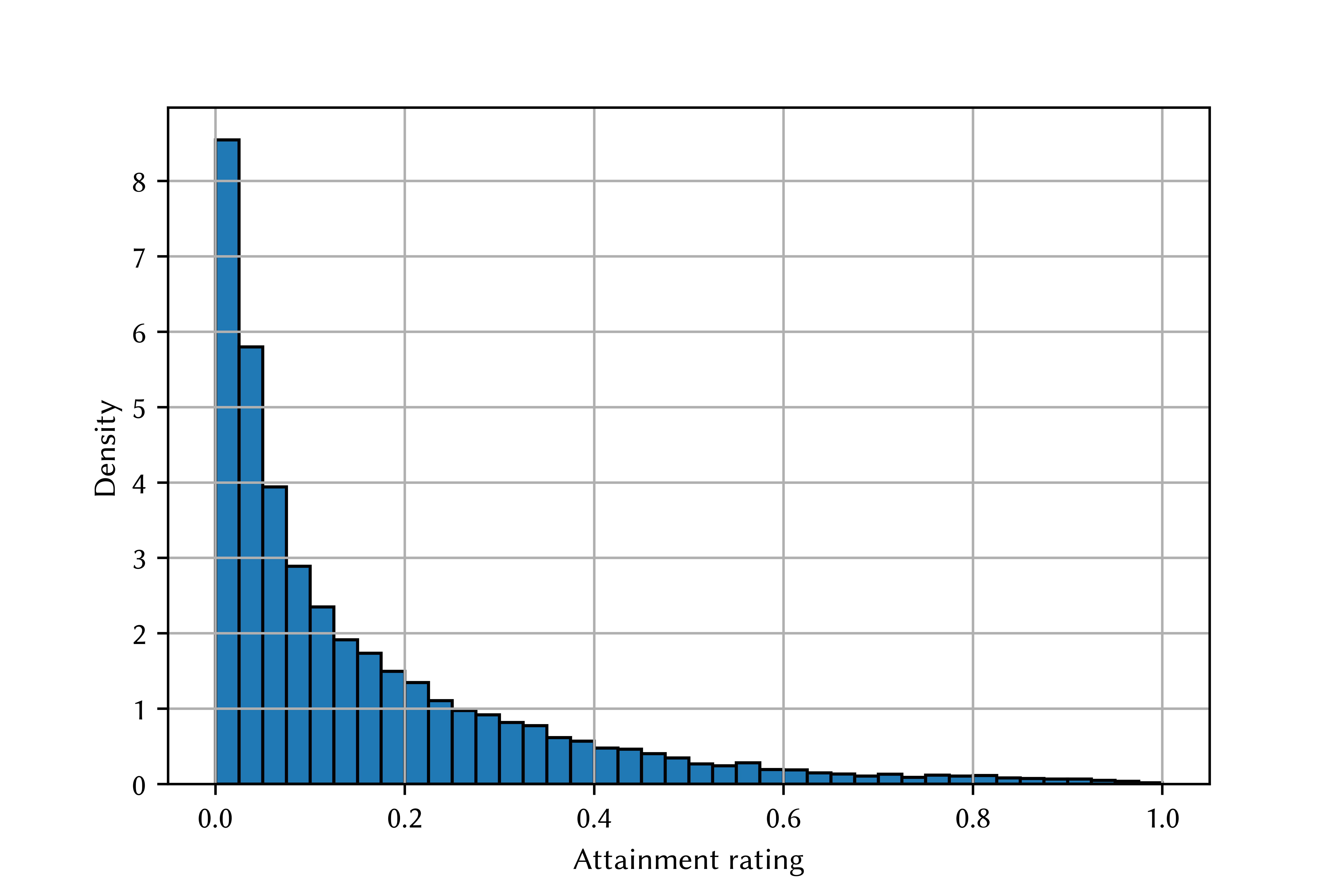}
	\caption{Distribution of attainment ratings across all users and games in crawled Steam dataset.}
	\label{fig:allattainments}
\end{figure}

We propose to eliminate these shortcomings by using achievement data to
define what we call an attainment rating. The attainment rating is
computed by combining each individual achievement
weighted by a difficulty level computed using 
% Through the use of examining the
% achievements that a particular user has gained for a game, we approximate
% the level of achievement. Furthermore, by using 
global achievement rates.  
For a game $g$ we denote its set
of achievements as $A_g=\{A_{g_1},A_{g_2},...,A_{g_{N_g}}\}$, where $N_g$
is the total number of achievements of game $g$. Each $A_{g_i} \in A_g$ is
a binary vector of length $\|P_g\|$, where $P_g$ is the set of players who
own game $g$, such that $A_{g_is}$ indicates whether or not player $s$ has
achieved achievement number $i$ in game $g$. Clearly, the proportion of
players who have achieved $A_{g_i}$ can be calculated as follows: 
\begin{equation}\label{eq:proportion_achieved}
C_{g_i}=\frac { \sum _{ p\in P_{ g } }^{  }{ A_{ g_{ i }p } }  }{ \| P_{ g }\|  }
\end{equation}
We define the attainment score $A_{gs}$ of player $s$ for game $g$ as follows:
\begin{equation}\label{eq:achievement_weighting}
A_{ gs }=\sum _{ A_{ g_{ i } }\in A_{ g } } \frac { A_{ g_{ i }s }\cdot
  \left( 1-C_{g_i} \right)  }{ { N_{ g } } }  
\end{equation}
The attainment score $ A_{gs} $ has a number of  desirable properties.
% nstance, Equation 
From~\eqref{eq:achievement_weighting}  it follows that that $A_{gs}$
aggregates a global
understanding of achievement difficulty -- a rare achievement,
i.e. one with $C_{g_i} \approx  0$,  makes a large contribution to
$A_{gs}$. Similarly, extremely common achievements, i.e. those with $C_{g_i}
\approx 1 $, contribute very little to the attainment
rating. From~\eqref{eq:achievement_weighting} we have:
\begin{equation}\label{eq:boundaries}
0\leq A_{ gs }\leq 1-\frac { 1 }{ \| P_{ g }\|  } 
\end{equation}
The attainment rating $A_{gs} = 0$ if the user has no achievements for game $g$, or in
the very unlikely event that everyone
else has all the achievements for it. The maximum value $A_{gs} = 1-\frac { 1 }{
  \| P_{ g }\|  }$ 
% The maximum value
is 
achieved in the unlikely event that user $s$ has all achievements for game $g$, and is the only user to
have all the achievements. % We therefore have a bounded
% score that does not suffer from the issues of traditional ratings
 In a 
traditional rating system, different users may have entirely different
standards for what constitutes a given score (e.g. giving out full marks
for ``I liked this'' vs full marks for ``this is literally my favorite
game of all time''). In contrast, the attainment ratings in
\eqref{eq:achievement_weighting} are precise, and measure the same degree
of  
attainment for every user. The use of achievements can give scores across the full range for
all users; we observe this in Figure \ref{fig:allattainments},
with truly high attainment ratings possible, but increasingly rare. 

\section{Graph Dataset}\label{sec:graph_dataset}
The Steam WebAPI gives access to a wide range of data, including data
relating to games, game ownership, friendships, group (e.g. gaming
``clans'' or social groups representing real-world groups) membership and
achievements. Although not explicitly stated, it is reasonable to
expect that the data is stored server-side in relational tables rather than
in a graph database due to telling limitations in the design of the API. In our
work, we propose that using a graph database and querying methods to store
and explore the Steam data is appropriate given the contained friendship
and group social networks, ownership and achievement data, and size of the
overall network. This paper does not detail computational performance or
explicit systems implementation of our graph-based recommendation system, with
these factors and their deployment in a graph computing system being a
current focus of our ongoing research. 

The Steam dataset has been analyzed previously, but (to the best
of the authors' knowledge) ours is the first work to use achievement data for
recommendations and focus on graph structure beyond basic social
networks. Some existing papers \cite{Becker2012,Blackburn2011} analyze
Steam from a network perspective, but focus on the growth of the friend
network, or the propagation of cheaters through interaction. These papers
predate the widespread use of achievements data in Steam, and even recent
work \cite{ONeill2016} that provides the crawled dataset online does not
include achievement data. 

Our crawled dataset is outlined in Table \ref{tbl:graph-summary}. We note
that this is merely a subset of the available data that we have chosen to
use (and that we were able to crawl before the privacy change of April 10,
2018) for recommendation in this preliminary work. 

 %\usepackage{booktabs}
\begin{table}[]
	\centering
	\caption{Vertex and edge types in dataset.}
	\label{tbl:graph-summary}
		\resizebox{\columnwidth}{!}{%
\begin{tabular}{@{}llll@{}}
	\toprule
	Name  & Type   & Description                                               & Count  \\ \midrule
	$V_P$ & Vertex & Players found in BFS of friend network of originating user     & 4159   \\
	$V_G$ & Vertex & Games. Contains content tags and structured summary data & 4487   \\
	$V_D$ & Vertex & Game developers (many ``indie'' developers making 1 game) & 1904   \\
	$V_R$ & Vertex & Game genres                                              & 30     \\
	$E_F$ & Edge   & Friendships between players $V_P$                        & 272888 \\
	$E_O$ & Edge   & Ownership by players in $V_P$ of games in $V_G$          & 613769 \\
	$E_D$ & Edge   & Multiple developers in $V_D$ developed games in $V_G$   & 4589   \\
	$E_R$ & Edge   & Games in $V_G$ may belong to multiple genres $V_R$      & 11229  \\ \bottomrule
\end{tabular}
}
\end{table}

\section{Graph-Focused Recommendation}\label{sec:graph_rec}
The use of graph-structured data in recommendation is a well developed
field \cite{Gori,Agichtein,Zhanga,Jamali2009}, with numerous techniques designed to incorporate network structural
information into recommendation systems. Such work has typically focused
on key domain areas such as web search \cite{Rogers2005,Fujiwara2012,Haveliwala2002} and friend recommendation in
social networks \cite{AlHasan2011,Liben-Nowell2007}. Network models (when used for recommendation) are often
fairly basic, with relatively few node and edge types (e.g. only having
``users'' and ``products'' as vertex types). The reasoning behind this
approach is quite simple; the presence of many different types of nodes
and edges can mean that the overabundance of a particular node or edge
type can dominate when applying traditional ranking  methodologies \cite{Aggarwal2016}.

%To demonstrate this phenomenon, consider the $E_D$ edges corresponding to
%game development. There are far fewer of these edges than there are
%friendships between players, meaning that a naive application of
%e.g. PageRank, would result in developer nodes and development edges
%having relatively small influence on the ranking of game products in the
%network in comparison to a users social network. 
%
%There have been a number of approaches developed to deal with this
%particular issue, such as those that perform rankings separately for each
%vertex type, and synthesize the results into a coherent overall ranking
%score. Our current research on probabilistic graph querying methodologies
%instead focuses on traversal mechanisms that are cognizant of node and
%edge types. 

In this work, we demonstrate a basic graph query approach for obtaining
flexible personalized recommendations using attainment ratings. Instead of
focusing on overall rankings, we demonstrate how the richness of the graph
structure can be used to obtain recommendations for top quality products
that match the users' specific requirements. We do not cast this as a complete solution to
the problem of video game recommendation as such, but rather use it as a
proof of concept to demonstrate the value of attainment ratings and
querying diverse graph-based datasets. 

The social aspect of using (attainment) rating data from actual friends (as opposed to
other users designated as ``similar'' via some clustering or similarity
measure) makes particular sense in the context of video games, where online
multiplayer 
is often a big part of a product's
appeal. This suggests the value of the graph-based recommendation in this
setting, where users can easily specify queries that account for desired
properties and social networks.  

We express example product search queries with SQL inspired
psuedo-code. For example, Listing \ref{lst:query1} is a graph query that
says ``for a certain user, find the names and purchase costs of five games
that are owned by friends, that are developed by companies who have
developed games already owned by the user, and order responses according
to how the friends `rated' them with respect to attainment''. In this
psuedo-code, graph objects are as defined in Table
\ref{tbl:graph-summary}. We search in the graph for specific paths of
nodes and edges \footnote{adding $(a)$ and $(b)$ etc. to the patterns
  identifies that these are the same ``objects'' in each pattern that must
  be satisfied}that can succinctly and interpretably describe ``complex''
queries that actually represent intuitive ideas. Though video games are
not in themselves a huge expense in comparison to purchases such as
houses, cars etc., we argue that their time investment \textit{is} large,
and that users are thereby willing to spend effort to be precise about
their requirements.

\begin{lstlisting}[caption={Sample graph query pseudo-code},escapeinside={(*}{*)},label={lst:query1}]
SELECT (*$V_G(b)$*).name, (*$V_G(b)$*).cost
PATTERNS (*$V_P(a)-E_F-V_P-E_O-V_G(b)$*)
(*$V_P(a)-E_O-V_G-E_D-V_D-E_D-V_G(b)$*)
WHERE (*$V_P(a)$*).steamid=76561197960653976
ORDERYBY AVG((*$V_P(a)-E_F-V_P-E_O.attainmentRating-V_G(b)$*))
LIMIT 5
\end{lstlisting}

\section{Experimental Results}\label{sec:results}
Here we demonstrate our preliminary experimental results. In this section,
we cover two main areas. The first is to summarize the crawl data as it
pertains to our derivation of attainment ratings (including the use of
attainment ratings in traditional collaborative filtering models). The
second is to demonstrate real-world applications of highly customizable
graph query recommendations. 

\subsection{The Value of Attainment Ratings}\label{subsec:attain_value}
The plots in Figure \ref{fig:allattainments} show that the distribution of
all crawled attainments is very regular, exhibiting similar properties to
a Lomax distribution (that is, a Pareto distribution shifted to begin at
zero) truncated at the maximum value (see
\eqref{eq:boundaries}). Indeed,a Kolmogorov-Smirnov test gives a very small
KS statistic ($KS=0.045$) when the ECDF is compared with a $Lomax(shape=4.78,
scale=0.61)$ distribution. The heavy tail of the empirical distribution
can be explained by the existence of ``achievement hunters'',
those who play in order to attain achievements, rather than the act of
playing being an end in and of itself. 

\begin{figure}
	\centering
	\includegraphics[width=1.0\linewidth]{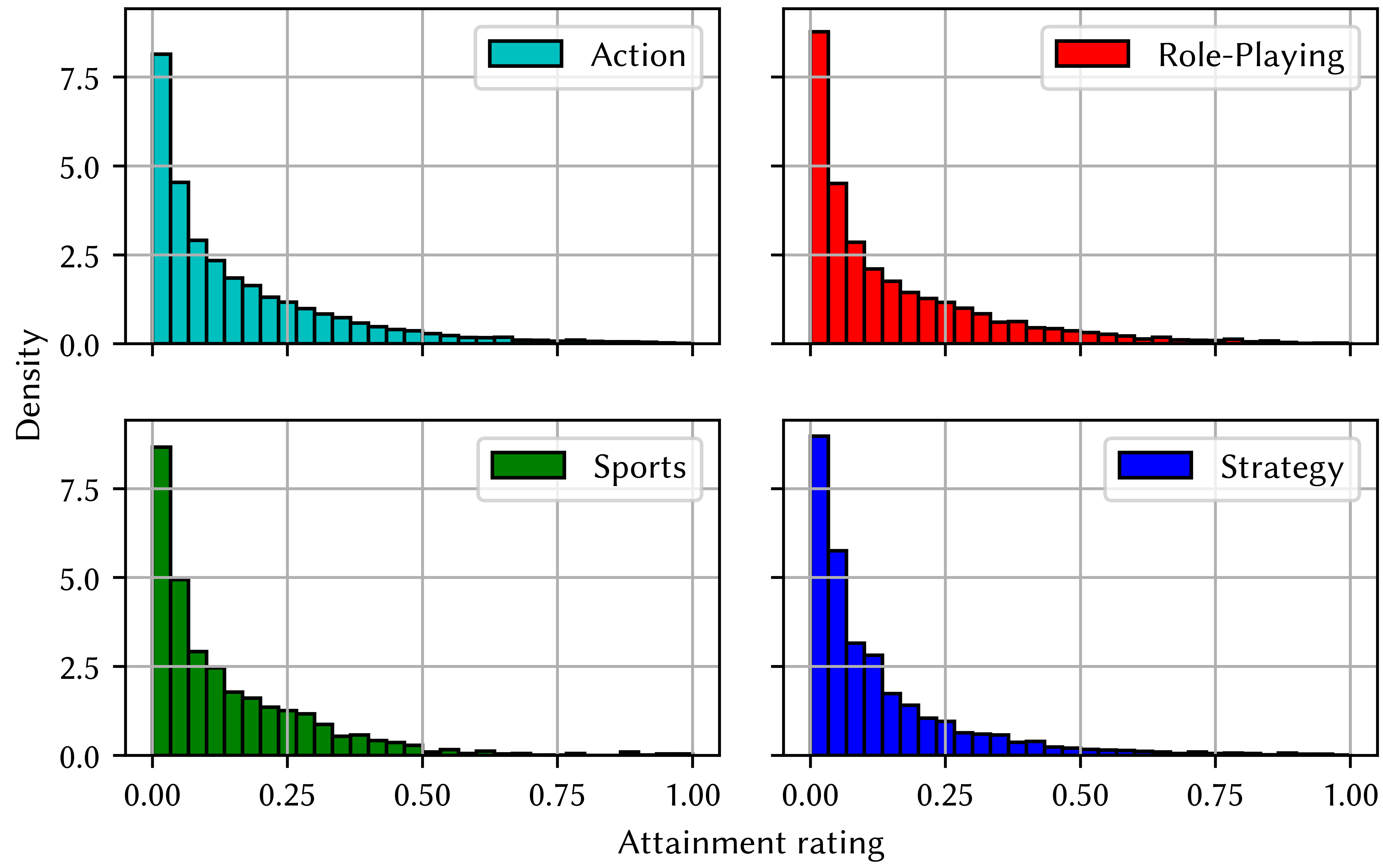}
	\caption{Density normalized histograms of attainment data grouped by example game genres.}
	\label{fig:genreattainmentshistogram}
\end{figure}

Of course, these results relate to the distribution of attainment
ratings across all combinations of users and their games. The distribution
of attainment ratings varies across developer, genre (see Figure
\ref{fig:genreattainmentshistogram}) etc. The differences in distribution
across these factors provide evidence of commonly held notions within the
video game community. For instance, we note that the tails of the Strategy
genre are far lighter than the Role-Playing genre, an observation
consistent with the idea that excelling in Strategy games requires
``talent'', whereas excelling in Role-Playing games is often a simple
matter of time investment. We also note that Action games have less mass
closer to zero, demonstrating their popularity and tendency for quick
satisfaction due to their dense but short nature. 

%\begin{figure}
%	\centering
%	\includegraphics[width=1.0\linewidth]{developer_attainments_histogram}
%	\caption{Density normalized histograms of attainment data grouped by example developers.}
%	\label{fig:developerattainmentshistogram}
%\end{figure}

\begin{figure}
	\centering
	\includegraphics[width=1.0\linewidth]{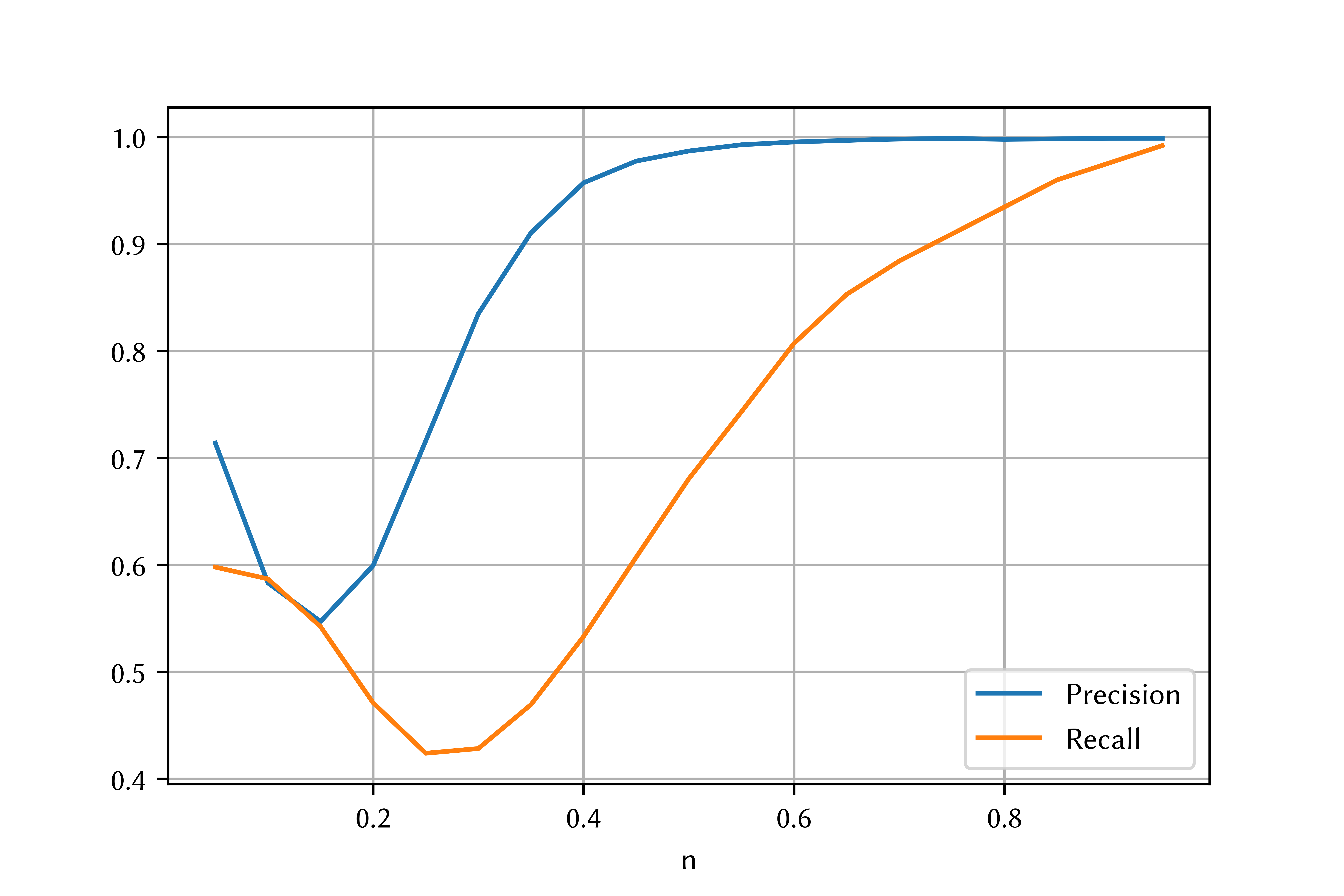}
	\caption{$Precision@n$ and $Recall@n$ for the range of attainment ratings.}
	\label{fig:precisionandrecall}
\end{figure}

Next, we demonstrate the use of attainment ratings in a
% recommendation systems by means of 
collaborative filtering (specifically,
SVD++ \cite{Student2014} due to the implicit nature of the attainment
ranking data) based recommendation system. Using the default parameters of the Surprise \cite{Hug2017}
implementation (e.g. 20 factors, 20 iterations of SGD with a learning rate
of 0.007) of SVD++, we obtained the results shown in Table
\ref{tbl:svdpp_err} and Figure \ref{fig:precisionandrecall}. It is clear
from these results that the attainment ratings
% have
% considerable value in their ability to 
well predict other attainment ratings
(as shown by Table \ref{tbl:svdpp_err}). We also observe from Figure
\ref{fig:precisionandrecall} that recommendations made using attainment
ratings result in highly relevant recommendations. We argue that the
somewhat lower recall values are to be expected from the data; buying many
games to be placed into a ``backlog'' before playing is a common practice. 

\begin{table}[]
	\centering
	\caption{RMSE and MAE using SVD++ with 5-fold cross-validation}
	\label{tbl:svdpp_err}
	\begin{tabular}{@{}llllll@{}}
		\toprule
		Fold & 1     & 2     & 3     & 4     & 5     \\ \midrule
		RMSE & 0.157 & 0.156 & 0.157 & 0.156 & 0.159 \\
		MAE  & 0.116 & 0.116 & 0.117 & 0.115 & 0.116 \\ \bottomrule
	\end{tabular}
\end{table}

\subsection{The Flexibility of Graph Query Recommendation}\label{subsec:graph_recs}
The average playtime of a video game is considerably longer than many
other forms of entertainment, with games in the Action genre having
playtimes typically in the 10-20 hour range, and Role-Playing games often
extending into \textit{hundreds} of hours.

Given the comparatively long length of video games and the large number of
products in the dataset, the ability to recall even a small fraction of
highly relevant products using attainment ratings suggests an ability to
generate a sufficiently large number of recommendations for users for the
duration of their use of the platform. This encourages the repeated use of
more personalized queries, where a user can narrow these general
recommendations to match their specific desires of the moment. 

We demonstrate this by means of running Listing \ref{lst:query1} for a
randomly chosen example user, along with some simple but useful
modifications. For instance, we can add the clause ``ANTIPATTERNS
$V_P(a)-E_O-V_G(b)$'' to encode that we do \textit{not} want to match
graph patterns that represent games that the user already owns. These are
represented as ``Sample Query'' and ``New Game Query'' respectively in
Table \ref{tbl:graph_recs}. Note that the attainment ratings as listed are
calculated with respect to the immediate neighborhood of the user in the
friend network, and not the entire user base. Not only are we explicitly indicating a preference to play games that friends like (and thereby play online \textit{with} them), trust-based work \cite{Golbeck2005} has shown that recommendations from more distant hops of the social network are less reliable. We observe that four of the
five games in Sample Query (all but ``Frozen Synapse'') are Action games,
with a single Strategy recommendation. When we perform New Game Query, this mix is also noted, with recommendations for both prominent Strategy games (``Magic:
The Gathering 2013'' and ``Total War: Warhammer'') and Action games
(``Call of Duty: Black Ops'' and ``Hitman: Absolution'').

This taste for Strategy games might be noted by the user, and it is a simple matter to restrict results to the Strategy genre (represented by ``Strategy Genre Query'' in Table \ref{tbl:graph_recs}) by adding the pattern $V_G(b)-E_R-V_R$ and the WHERE clause condition $V_R.description=Strategy$.
\begin{table}[]
	\centering
	\caption{Example graph query recommendations}
	\label{tbl:graph_recs}
	\resizebox{1.0\columnwidth}{!}{%
\begin{tabular}{@{}ccccccc@{}}
	\toprule
	\multicolumn{1}{l}{} & \multicolumn{2}{c}{Sample Query} & \multicolumn{2}{c}{New Game Query}     & \multicolumn{2}{c}{Strategy Genre Query}       \\ \midrule
	Rank                 & Name              & Attainment   & Name                      & Attainment & Name                              & Attainment \\
	1                    & APB Reloaded      & 0.646        & Magic: The Gathering 2013 & 0.233      & Magic: The Gathering 2013         & 0.233      \\
	2                    & Warframe          & 0.516        & Total War: Warhammer      & 0.198      & Total War: Warhammer              & 0.198      \\
	3                    & Left 4 Dead       & 0.407        & Call of Duty: Black Ops   & 0.161      & Warhammer 40,000: Dawn of War III & 0.076      \\
	4                    & Team Fortress 2   & 0.347        & Hitman: Absolution        & 0.159      & Age of Empires II HD              & 0.018      \\
	5                    & Frozen Synapse    & 0.292        & Super Star                & 0.109      & Fray: Reloaded Edition            & 0.013      \\ \bottomrule
\end{tabular}}
\end{table}

\section{Future Work}\label{sec:conclusion}
With Steam data now effectively private, we are crawling Xbox Live to
serve as a more complete dataset than is possible with our Steam
data, allowing us to make increasingly general claims, and improve our measures of performance.

Furthermore, we are developing a graph querying methodology that allows
for the efficient specification and implementation of customizable graph
queries, such as those that were outlined in Section
\ref{subsec:graph_recs}. This graph querying methodology, based on
probabilistic traversals of the graph, is intended to facilitate a
customizable video game recommendation solution that is capable of scaling
to the enormous sizes of the social, purchase, and attainment graphs
associated with online videogame marketplaces like Steam and Xbox Live. 

This paper has been a practical demonstration of the wealth of marketable information contained
within achievement data. We also seek to gain a more theoretical understanding of how to best 
employ this data for the purpose of recommendation.

\bibliographystyle{ACM-Reference-Format}
\bibliography{sample-sigconf}

\end{document}